\documentclass[12pt]{iopart}
\renewcommand{\theequation}{\Roman{section}.\arabic{equation}}
\usepackage{graphicx}

\begin{document}

\title{Efficiency of Carnot Cycle with Arbitrary Gas Equation of
State}
\author{Paulus C.\ Tjiang$^1$ and
Sylvia H.\ Sutanto$^2$}
\address{Department of Physics, Faculty of Mathematics and Natural Sciences \\
Universitas Katolik Parahyangan, Bandung 40141 - INDONESIA}

\ead{$^1$pctjiang@home.unpar.ac.id, $^2$sylvia@home.unpar.ac.id}

\begin{abstract}
The derivation of the efficiency of Carnot cycle is usually done by
calculating the heats involved in two isothermal processes and
making use of the associated adiabatic relation for a given working
substance's equation of state, usually the ideal gas. We present a
derivation of Carnot efficiency using the same procedure with
Redlich-Kwong gas as working substance to answer the calculation
difficulties raised by Agrawal and Menon~\cite{AM90}. We also show
that using the same procedure, the Carnot efficiency may be derived
regardless of the functional form of the gas equation of state.

\end{abstract}


\section{Introduction.}
\label{intro}

In any course of undergraduate thermodynamics, thermodynamical
cycles and their efficiencies hardly miss their parts. The
discussion of thermodynamical cycles is always followed by, or is
parallel to, the discussion of the second law of thermodynamics. For
a reversible cycle between two single temperatures, it is well-known
that the efficiency $\eta$ of such cycle is
\begin{equation}
\eta=1- \frac{T_C}{T_H}, \label{efficiency}
\end{equation}
where $T_H$ and $T_C$ are the absolute temperatures of hot and cold
heat reservoirs. For an irreversible cycle, since the total change
of entropy is positive, the efficiency of such cycle is less than
(\ref{efficiency}).

\begin{figure}
\begin{center}
\includegraphics[scale=0.5]{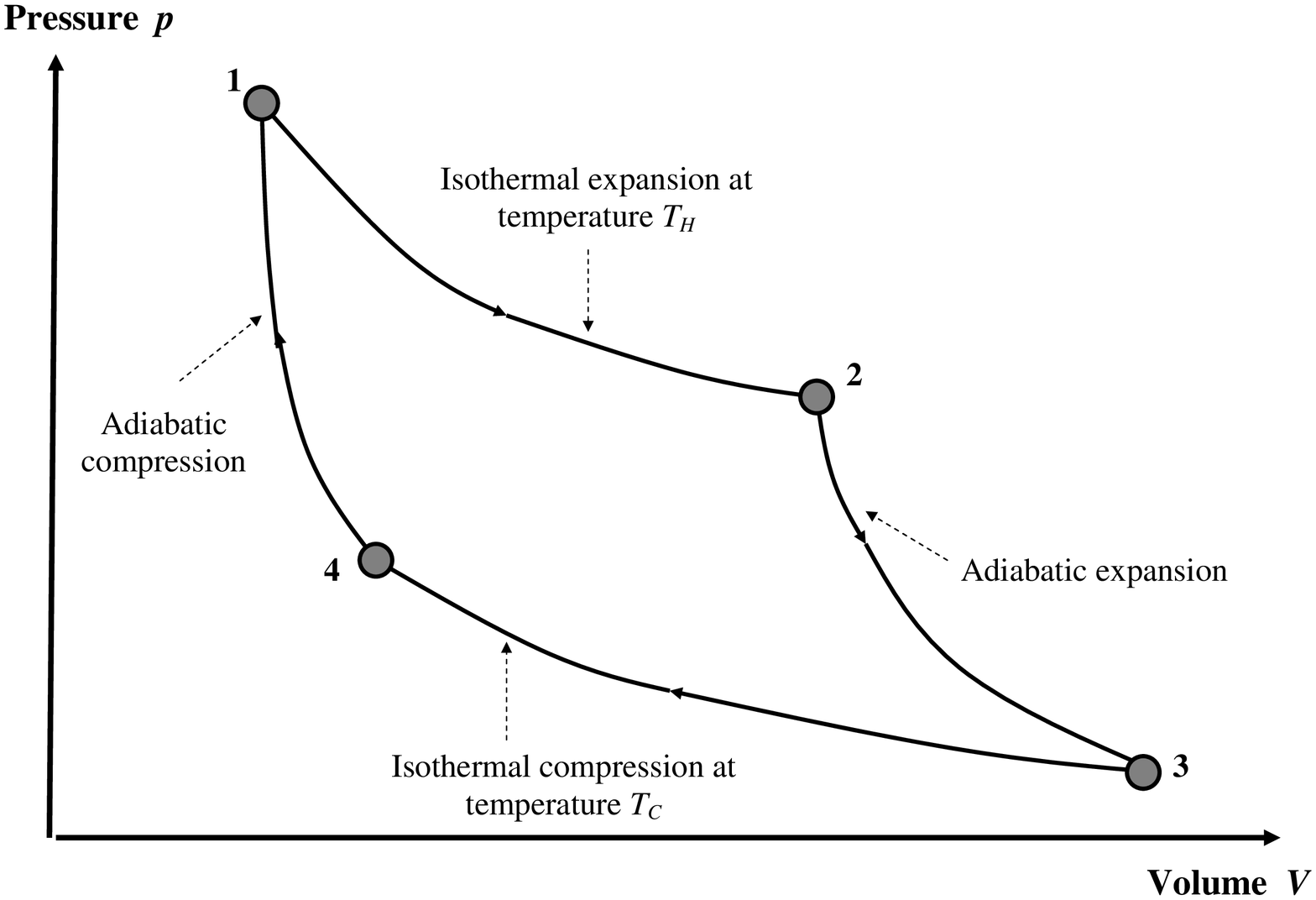}
\caption{The $p - V$ indicator diagram of a Carnot cycle 1-2-3-4,
where $T_C < T_H$.} \label{Carnot}
\end{center}
\end{figure}

There are many theoretical cycles that satisfy the efficiency
(\ref{efficiency})~\cite{KW77}, but the so-called {\it Carnot cycle}
is of interest because of its simple manner in describing a
reversible cycle. The Carnot cycle consists of two isothermal
processes and two adiabatic processes as shown in \Fref{Carnot}. In
most textbooks of either elementary physics or thermodynamics, the
Carnot efficiency (\ref{efficiency}) is derived with the ideal gas
as its working substance because of its mathematical simplicity
using the following customary procedure : {\it calculating the heats
involved in isothermal expansion and compression in $p - V$, $V - T$
or $p - T$ diagrams, then relating them with the associated
adiabatic relations}. However, the second law of thermodynamics
proves that the efficiency (\ref{efficiency}) should be independent
of working substance~\cite{Sears75}, so it is natural to question
whether the Carnot efficiency can be obtained from other equations
of state using the procedure above. Some attempt has been made to do
the task above, among them is the work of Agrawal and
Menon~\cite{AM90} who used the van der Waals equation of state to
derive the Carnot efficiency through the procedure above, and it
turned out that their result agreed with (\ref{efficiency}).
Nevertheless, they pointed out that there were some calculation
difficulties arising when the derivation was done for other real
gases' equations of state (such as the Redlich-Kwong
gas~\cite{KW77}) using the same procedure, and suggested to derive
Eq.(\ref{efficiency}) through an infinitesimal cycle using a
suitable Taylor expansion of thermodynamic variables about the
initial state of the cycle~\cite{AM90}.

In this paper we shall derive the Carnot efficiency
(\ref{efficiency}) with the Redlich-Kwong gas as working substance
to answer the calculation difficulties raised by Agrawal and Menon,
and we also show that using the customary procedure, we may obtain
the efficiency (\ref{efficiency}) regardless of the functional form
of the equation of state. We start with a brief review of the
generalized thermodynamic properties satisfied by any equation of
state in Section~\ref{general-property}. Using the relations
discussed in Section~\ref{general-property}, we shall derive the
Carnot efficiency from the Redlich-Kwong equation of state in
Section~\ref{redlich-kwong}. In Section~\ref{arbitrary}, we present
the derivation of the Carnot efficiency without any knowledge of the
substance's equation of state using the customary procedure. The
discussion will be concluded in Section~\ref{summary}.

\section{Generalized Thermodynamic Properties}
\label{general-property}

In this section we shall briefly review some thermodynamic
properties satisfied by any equation of state.

\subsection{Maxwell Relations}
\label{maxwell-relation}

An equation of state in thermodynamics may be written as
\begin{equation}
f(p,V,T) = C, \label{state-function}
\end{equation}
where $p$, $V$ and $T$ are pressure, volume and absolute
temperature of substance, respectively. Eq.~(\ref{state-function})
yields the following relations :
\begin{eqnarray}
p & = & p(V,T), \label{p-function} \\
V & = & V(p,T), \label{V-function} \\
T & = & T(p,V). \label{T-function}
\end{eqnarray}
However, the first law of thermodynamics and the definition of
entropy suggests that there is another degree of freedom that
should be taken into account, i.e. the entropy $S$, for
\begin{equation}
dU = T dS - p dV \longrightarrow U = U(S,V), \label{U-SV}
\end{equation}
where $U$ is the internal energy of the substance. From
Eq.~(\ref{U-SV}), it is clear that
\begin{eqnarray}
\left(\frac{\partial U}{\partial S} \right)_V & = & T, \nonumber \\
\left(\frac{\partial U}{\partial V} \right)_S & = & -p,
\end{eqnarray}
which gives
\begin{equation}
\left(\frac{\partial T}{\partial V} \right)_S = -
\left(\frac{\partial p}{\partial S} \right)_V
\label{Maxwell-relation-1}
\end{equation}
according to the exactness condition of the internal energy $U$.

Using Legendre transformations~\cite{Goldstein80}, we may define
\begin{eqnarray}
H(p,S) & = & U(S,V) + p V, \label{enthalpy} \\
F(V,T) & = & U(S,V) - T S, \label{Helmholtz} \\
G(P,T) & = & H(p,S) - T S, \label{Gibbs}
\end{eqnarray}
where $H(p,S)$, $F(V,T)$ and $G(p,T)$ are enthalpy, Helmholtz and
Gibbs functions, respectively. Differentiating
Eqs.~(\ref{enthalpy}), (\ref{Helmholtz}) and (\ref{Gibbs}) give
\begin{eqnarray}
dH & = & T dS + V dp, \\
dF & = & - p dV - S dT, \\
dG & = & V dp - S dT,
\end{eqnarray}
which lead us to
\begin{eqnarray}
\left(\frac{\partial T}{\partial p} \right)_S & = &
\left(\frac{\partial V}{\partial S} \right)_p,
\label{Maxwell-relation-2} \\
\left(\frac{\partial p}{\partial T} \right)_V & = &
\left(\frac{\partial S}{\partial V} \right)_T,
\label{Maxwell-relation-3} \\
\left(\frac{\partial V}{\partial T} \right)_p & = & -
\left(\frac{\partial S}{\partial p} \right)_T
\label{Maxwell-relation-4}
\end{eqnarray}
due to the exactness of $H(p,S)$, $F(V,T)$ and $G(p,T)$. The set of
Eqs.~(\ref{Maxwell-relation-1}), (\ref{Maxwell-relation-2}),
(\ref{Maxwell-relation-3}) and (\ref{Maxwell-relation-4}) is called
the Maxwell relations~\cite{KW77,Sears75}.

\subsection{General Properties of Entropy and Internal Energy}
\label{general-S-U}

Now let us express the entropy $U$ and internal energy $S$ in terms
of measurable quantities. Let $U = U(V,T)$, then
\begin{eqnarray}
dU & = & \left(\frac{\partial U}{\partial T}\right)_V dT +
\left(\frac{\partial U}{\partial V}\right)_T dV \nonumber \\
  & = & C_v dT + \left(\frac{\partial U}{\partial V}\right)_T dV,
\label{differential-U}
\end{eqnarray}
where $C_v$ is the heat capacity at constant volume. Inserting
Eq.~(\ref{differential-U}) into Eq.~(\ref{U-SV}), we have
\begin{equation}
dS = \frac{C_v}{T} dT + \frac{1}{T} \left[\left(\frac{\partial
U}{\partial V} \right)_T + p \right] dV. \label{differential-S-2}
\end{equation}
Suppose $S = S(T,V)$, then
\begin{eqnarray}
dS & = & \left(\frac{\partial S}{\partial T}\right)_V dT +
\left(\frac{\partial S}{\partial V}\right)_T dV. \nonumber \\
   & = & \left(\frac{\partial S}{\partial T}\right)_V dT +
\left(\frac{\partial p}{\partial T}\right)_V dV.
\label{differential-S}
\end{eqnarray}
where we have used Eq.~(\ref{Maxwell-relation-3}). Comparing
Eqs.~(\ref{differential-S}) and (\ref{differential-S-2}), we obtain
\begin{eqnarray}
\left(\frac{\partial S}{\partial T}\right)_V & = & \frac{C_v}{T}, \\
\left(\frac{\partial U}{\partial V}\right)_T & = & T
\left(\frac{\partial p}{\partial T} \right)_V - p.
\label{differential-S-3}
\end{eqnarray}
Substitution of Eq.~(\ref{differential-S-3}) into Eq.~(\ref{U-SV})
gives
\begin{equation}
dU = C_v dT  + \left[T \left(\frac{\partial p}{\partial T}
\right)_V - p \right] dV. \label{differential-U-2}
\end{equation}
Since $dU$ is an exact differential, the following exactness
condition must be fulfilled :
\begin{equation}
\left(\frac{\partial C_v}{\partial V}\right)_T = T
\left(\frac{\partial^2 p}{\partial T^2}\right)_V.
\label{exactness-U}
\end{equation}
It is easy to see that Eq.~(\ref{exactness-U}) must also be
satisfied to ensure the exactness of Eq.~(\ref{differential-S}).
Eq.~(\ref{exactness-U}) also tells us the isothermal volume
dependence of $C_v$.

\subsection{General Relations of Isothermal and Adiabatic Processes}
\label{general-isothermal-adiabatic}

In an {\it isothermal} process, the change of internal energy is
given by
\begin{equation}
dU = \left[T \left(\frac{\partial p}{\partial T} \right)_V - p
\right] dV, \label{dU-isothermal}
\end{equation}
using Eq.~(\ref{differential-U-2}). Using the first law of
thermodynamics $dU = dQ - p \ dV$, the heat involved in this process
is
\begin{equation}
dQ = T \left(\frac{\partial p}{\partial T} \right)_V dV.
\label{dQ-isothermal}
\end{equation}

In an {\it adiabatic} process where no heat is involved, the first
law of thermodynamics, together with Eq.~(\ref{differential-U-2})
gives
\begin{equation}
C_{v} dT = -T \left(\frac{\partial p}{\partial T} \right)_V dV
\label{general-adiabatic}
\end{equation}

Equations (\ref{dQ-isothermal}) and (\ref{general-adiabatic}) will
be used to obtain the Carnot efficiency of the Redlich-Kwong gas in
the next section.

\section{Carnot Efficiency of the Redlich-Kwong Equation of State}
\label{redlich-kwong}

In this section we shall derive the Carnot efficiency
(\ref{efficiency}) from the Redlich-Kwong gas, whose equation of
state is given by
\begin{equation}
p = \frac{n R T}{V - b} - \frac{n^2 a}{T^{1/2} V (V+b)}, \label{R-K}
\end{equation}
where $n$ is the number of moles of the gas, $R \approx 8.31 \ J \
mol^{-1} K^{-1}$ is the gas constant, $a$ and $b$ are constants
evaluated from the critical state of the gas~\cite{KW77}. We shall
follow the process order of the Carnot cycle as shown in
\Fref{Carnot}.

From Eq.~(\ref{exactness-U}), the volume dependence of the heat
capacity of constant volume $C_v$ for the Redlich-Kwong gas is
\begin{equation}
\left(\frac{\partial C_v}{\partial V}\right)_T = - \frac{3 n^2 a}{4
T^{3/2} V (V+b)},
\end{equation}
which leads to the following functional form of $C_v$ :
\begin{equation}
C_v (V,T) = \frac{3 n^2 a}{4 b T^{3/2}} \ln{\frac{V + b}{V}} + f(T),
\label{Cv-RK}
\end{equation}
where $f(T)$ is an arbitrary function of temperature, since we do
not have any information of $\left(\frac{\partial C_v}{\partial
T}\right)_V$.

Using Eqs.~(\ref{dQ-isothermal}) and (\ref{R-K}), we obtain the
involved heat for the isothermal expansion from states 1 to 2, as
well as isothermal compression from states 3 to 4 as follows :
\begin{eqnarray}
Q_{1 \rightarrow 2} & = & n R T_H \ln \frac{V_2 - b}{V_1 - b} +
\frac{n^2 a}{2 b T_H^{1/2}} \ln \frac{V_2 (V_1 + b)}{V_1 (V_2 + b)},
\label{isothermal-H} \\
Q_{3 \rightarrow 4} & = & n R T_C \ln \frac{V_4 - b}{V_3 - b} +
\frac{n^2 a}{2 b T_C^{1/2}} \ln \frac{V_4 (V_3 + b)}{V_3 (V_4 + b)}.
\label{isothermal-C}
\end{eqnarray}

For the adiabatic process, Eq.~(\ref{general-adiabatic}) leads to
the following differential form with the help of Eq.~(\ref{Cv-RK}) :
\begin{eqnarray}
M(V,T) \ dT & + &  N(V,T) \ dV = 0, \label{adiabatic-RK-DE} \\
M(V,T) & = & \frac{3 n^2 a}{4 b T^{3/2}} \ln{\frac{V + b}{V}} +
f(T), \label{M}
\\
N(V,T) & = & \frac{n R T}{V - b} + \frac{n^2 a}{2 T^{1/2} V (V+b)}.
\label{N}
\end{eqnarray}
It is clear that Eq.~(\ref{adiabatic-RK-DE}) is not an exact
differential, which means that we have to find a suitable
integrating factor in order to transform Eq.~(\ref{adiabatic-RK-DE})
to an exact differential. The correspondence integrating factor
$\mu(V,T)$ for Eq.~(\ref{adiabatic-RK-DE}) is surprisingly simple :
\begin{equation}
\mu(V,T) \longrightarrow \mu(T) = \frac{1}{T}.
\label{integrating-factor}
\end{equation}
Multiplying $\mu(T)$ to Eq.~(\ref{adiabatic-RK-DE}) gives
\begin{eqnarray}
\bar{M}(V,T) \ dT & + &  \bar{N}(V,T) \ dV = 0, \label{adiabatic-RK-exact} \\
\bar{M}(V,T) & = & \frac{3 n^2 a}{4 b T^{5/2}} \ln{\frac{V + b}{V}}
+ \frac{f(T)}{T}, \label{bar-M}
\\
\bar{N}(V,T) & = & \frac{n R}{V - b} + \frac{n^2 a}{2 T^{3/2} V
(V+b)}, \label{bar-N}
\end{eqnarray}
whose general solution is
\begin{equation}
n R \ln (V-b) + \frac{n^2 a}{2 b T^{3/2}} \ln \frac{V}{V+b} + g(T) =
\mbox{constant}, \label{adiabatic-solution}
\end{equation}
where
\begin{equation}
g(T) = \int \frac{f(T)}{T} \ dT.
\end{equation}

Using Eq.~(\ref{adiabatic-solution}), we obtain the relation between
states $2$ and $3$ connected by adiabatic expansion as
\begin{eqnarray}
&  & n R \ln (V_2-b) + \frac{n^2 a}{2 b T_H^{3/2}} \ln
\frac{V_2}{V_2+b} + g(T_H) \nonumber \\
& = & n R \ln (V_3-b) + \frac{n^2 a}{2 b T_C^{3/2}} \ln
\frac{V_3}{V_3+b} + g(T_C). \label{adiabatic-BC}
\end{eqnarray}
The similar relation holds for adiabatic compression from states $4$
to $1$ :
\begin{eqnarray}
&  & n R \ln (V_1-b) + \frac{n^2 a}{2 b T_H^{3/2}} \ln
\frac{V_1}{V_1+b} + g(T_H) \nonumber \\
& = & n R \ln (V_4-b) + \frac{n^2 a}{2 b T_C^{3/2}} \ln
\frac{V_4}{V_4+b} + g(T_C). \label{adiabatic-DA}
\end{eqnarray}
Eqs.~(\ref{adiabatic-BC}) and (\ref{adiabatic-DA}) may be rewritten
as
\begin{eqnarray}
g(T_H) - g(T_C) & = & n R \ln \frac{V_3-b}{V_2-b} + \frac{n^2 a}{2 b
T_C^{3/2}} \ln \frac{V_3}{V_3+b} \nonumber \\
& - & \frac{n^2 a}{2 b T_H^{3/2}} \ln \frac{V_2}{V_2+b}
\label{adiabatic-BC-1}
\end{eqnarray}
and
\begin{eqnarray}
g(T_H) - g(T_C) & = & n R \ln \frac{V_4-b}{V_1-b} + \frac{n^2 a}{2 b
T_C^{3/2}} \ln \frac{V_4}{V_4+b} \nonumber \\
& - & \frac{n^2 a}{2 b T_H^{3/2}} \ln \frac{V_1}{V_1+b},
\label{adiabatic-DA-1}
\end{eqnarray}
respectively. Equating Eqs.~(\ref{adiabatic-BC-1}) and
(\ref{adiabatic-DA-1}) and after doing some algebraic calculation,
we get
\begin{eqnarray}
&   & n R \ln \frac{V_2 - b}{V_1 - b} + \frac{n^2 a}{2 b T_H^{3/2}}
\ln \frac{V_2 (V_1+b)}{V_1 (V_2+b)} \nonumber \\
& = & n R \ln \frac{V_3 - b}{V_4 - b} + \frac{n^2 a}{2 b T_C^{3/2}}
\ln \frac{V_3 (V_4+b)}{V_4 (V_3+b)}. \label{adiabatic-RK-relation}
\end{eqnarray}

Now let us calculate the Carnot efficiency of Redlich-Kwong gas.
From Eqs.~(\ref{isothermal-H}) and (\ref{isothermal-C}), the
efficiency $\eta$ is
\begin{eqnarray}
\eta & = & \frac{|Q_{1 \rightarrow 2}| - |Q_{3 \rightarrow 4}|}{|Q_{1 \rightarrow 2}|} = 1 - \frac{|Q_{3 \rightarrow 4}|}{|Q_{1 \rightarrow 2}|} \nonumber \\
     & = & 1 - \frac{T_C \left(n R \ln \frac{V_3 - b}{V_4 - b} +
\frac{n^2 a}{2 b T_C^{3/2}} \ln \frac{V_3 (V_4 + b)}{V_4 (V_3 +
b)}\right)}{T_H \left(n R \ln \frac{V_2 - b}{V_1 - b} + \frac{n^2
a}{2 b T_H^{3/2}} \ln \frac{V_2 (V_1 + b)}{V_1 (V_2 + b)}\right)}
\longrightarrow 1 - \frac{T_C}{T_H} \label{Carnot-efficiency-RK}
\end{eqnarray}
where we have used the adiabatic relation
(\ref{adiabatic-RK-relation}). It is clear that the Carnot
efficiency (\ref{Carnot-efficiency-RK}) coincides with
Eq.~(\ref{efficiency}) in the Section~\ref{intro} of this paper.

\section{Derivation of Carnot Efficiency of Arbitrary Gas Equation of State}
\label{arbitrary}

The success of obtaining Carnot efficiency with the van der Waals
gas in Ref.~\cite{AM90} and the Redlich-Kwong gas in the previous
section has tempted us to question whether we may obtain
Eq.~(\ref{efficiency}) from any working substance using the same
procedure mentioned in Section~\ref{intro}. Let the substance's
equation of state be in the form of Eq.~(\ref{p-function}). With the
volume dependence of $C_v$ is given by Eq.~(\ref{exactness-U}), the
functional form of $C_v$ is
\begin{equation}
C_v (V,T) = T \int \left(\frac{\partial^2 p}{\partial T^2}\right)_V
dV + f (T), \label{Cv-general}
\end{equation}
where $f(T)$ is an arbitrary function of temperature.

Using the same process order of Carnot cycle as given in
\Fref{Carnot} and with help of Eq.~(\ref{dQ-isothermal}), the
involved heat in isothermal expansion from states $1$ to $2$, as
well as isothermal compression from states $3$ to $4$ are
\begin{eqnarray}
Q_{1 \rightarrow 2} & = & T_H \int_{V_1}^{V_2} \
\left(\frac{\partial p}{\partial T} \right)_V \ dV \equiv T_H \left[F(V_2,T_H) - F(V_1,T_H) \right], \label{heat-general-1-2}\\
Q_{3 \rightarrow 4} & = & T_C \int_{V_3}^{V_4} \
\left(\frac{\partial p}{\partial T} \right)_V \ dV \equiv T_C
\left[F(V_4,T_C) - F(V_3,T_C) \right], \label{heat-general-3-4}
\end{eqnarray}
respectively, where $F(V,T) = \int \left(\frac{\partial p}{\partial
T} \right)_V \ dV$.

In the adiabatic process, with the help of Eq.~(\ref{exactness-U})
it is easy to see that Eq.~(\ref{general-adiabatic}) is not an exact
differential. However, by multiplying Eq.~(\ref{general-adiabatic})
with a suitable integrating factor, which turns out to be $\mu(V,T)
= \frac{1}{T}$ like the one used in Section~\ref{redlich-kwong}, we
obtain
\begin{equation}
\frac{C_{v}}{T} \ dT + \left(\frac{\partial p}{\partial T} \right)_V
dV = 0. \label{exact-adiabatic}
\end{equation}
With the help of Eq.~(\ref{Cv-general}), it is easy to see that
Eq.~(\ref{exact-adiabatic}) is an exact differential, whose general
solution is
\begin{equation}
\int \left(\frac{\partial p}{\partial T} \right)_V dV + g(T) =
\mbox{constant} \longrightarrow F(V,T) + g(T) = \mbox{constant},
\label{general-adiabatic-arbitrary}
\end{equation}
where $g(T) = \int \frac{f(T)}{T} \ dT$ is another arbitrary
function of temperature. Using
Eq.~(\ref{general-adiabatic-arbitrary}), the relation between states
$2$ and $3$ in the adiabatic expansion, as well as the relation
between states $4$ and $1$ in the adiabatic compression are
\begin{eqnarray}
g(T_H) - g(T_C) & = & F(V_3,T_C) - F(V_2,T_H),
\label{adiabatic-BC-relation-arbitrary} \\
g(T_H) - g(T_C) & = & F(V_4,T_C) - F(V_1,T_H),
\label{adiabatic-DA-relation-arbitrary}
\end{eqnarray}
respectively. Equating Eqs.~(\ref{adiabatic-BC-relation-arbitrary})
and (\ref{adiabatic-DA-relation-arbitrary}), we get
\begin{equation}
F(V_3,T_C) - F(V_4,T_C) = F(V_2,T_H) - F(V_1,T_H).
\label{adiabatic-ABCD-relation}
\end{equation}

Finally, the Carnot efficiency $\eta$ is
\begin{eqnarray}
\eta & = & 1 - \frac{|Q_{3 \rightarrow 4}|}{|Q_{1 \rightarrow 2}|} \nonumber \\
     & = & 1 - \frac{T_C
\left|F(V_4,T_C) - F(V_3,T_C) \right|}{T_H \left|F(V_2,T_H) -
F(V_1,T_H) \right|} \longrightarrow 1 - \frac{T_C}{T_H}
\end{eqnarray}
using Eq.~(\ref{adiabatic-ABCD-relation}). It is just the same
efficiency as Eq.~(\ref{efficiency}) given in Section~\ref{intro}.

\section{Summary and Conclusion}
\label{summary}

In this paper, we have derived the Carnot efficiency for the
Redlich-Kwong gas as well as for arbitrary gas equations of state
using the procedure given in Section~\ref{intro}. Both results are
in agreement with Eq.~(\ref{efficiency}).

From the derivation using the Redlich-Kwong gas equation of state,
we show that the derivation procedure succeeds even if the specific
heat at constant volume $C_v$ is the function of volume and
temperature - the difficulty encountered by Agrawal and
Menon~\cite{AM90} while deriving Carnot efficiency using equation of
state with $\left(\frac{\partial C_v}{\partial V} \right)_T \neq 0$.
As shown by Eq.~(\ref{Cv-RK}), we may write the analytical form of
$C_v (V,T)$ with an unknown function of temperature in it since we
know only the volume dependence of $C_v$ through
$\left(\frac{\partial C_v}{\partial V} \right)_T$. From
Eq.~(\ref{adiabatic-RK-relation}), it is clear that the equation of
adiabatic relations between states 1, 2, 3 and 4 does not depend on
that unknown function of temperature.

On the contrary of Agrawal-Menon's discussion in Ref.~\cite{AM90}
that it is difficult to apply the procedure stated in
Section~\ref{intro} for a finite Carnot cycle when the working
substance is arbitrary, our results in Section~\ref{arbitrary} show
that it is technically possible to derive the Carnot
efficiency~(\ref{efficiency}) from the general thermodynamic
properties discussed in Section~\ref{general-property}. However,
since the thermodynamic properties are derived from the Maxwell's
relations where the concept of entropy is used, the results in
Section~\ref{arbitrary} are hence not surprising. Using
Eqs.~(\ref{general-adiabatic}), (\ref{heat-general-1-2}) -
(\ref{heat-general-3-4}) and (\ref{adiabatic-ABCD-relation}), it is
easy to verify that the derivation given in Section~\ref{arbitrary}
is completely equivalent to the condition of a reversible cycle
$\oint dS = 0$ which also produces the Carnot
efficiency~(\ref{efficiency}) regardless of working substance. The
results in Section~\ref{arbitrary} may answer student's questions
concerning how the derivation of Carnot efficiency from any given
working substance may be carried out using the procedure stated in
Section~\ref{intro} to produce Eq.~(\ref{efficiency}) .

\section*{Acknowledgement}
The authors would like to thank Prof. B. Suprapto Brotosiswojo and
Dr. A. Rusli of the Department of Physics, Institut Teknologi
Bandung - Indonesia for their helpful discussions and corrections on
the subject in this paper.

\end{document}